





\magnification\magstep2
  \baselineskip = 0.65 true cm
  \parskip=0.09 true cm
                           
  \def\sa{\vskip 0.30 true cm}
  \def\sb{\vskip 0.60 true cm}

\pageno = 0 


\def\grt{\hbox{{\bf R}$^3$}}
\def\grq{\hbox{{\bf R}$^4$}}

\def\grn{\hbox{{\bf N}}}

\rightline{\bf LYCEN 9152}
\rightline{November 1991}

\sa
\sb
\sa

\noindent {\bf A $q$-deformed Aufbau Prinzip}

\sa
\sb

\noindent T N\'egadi\footnote*{Permanent address:  
Laboratoire de Physique Th\'eorique, 
Institut de Physique, Universit\'e d'Oran, 
Es-S\'enia, Oran, Alg\'erie} and M Kibler

\sa

\noindent Institut de Physique Nucl\'eaire de Lyon, 
IN2P3-CNRS et Universit\'e Claude Bernard, 
F-69622 Villeurbanne Cedex, France 

\sa
\sa
\sb
\sb


\sa

\noindent {\bf Abstract}. A building principle working for both 
atoms and monoatomic ions is proposed in this Letter. This 
principle relies on the $q$-deformed ``chain'' $SO(4) > SO(3)_q$. 

\sa
\sb
\sa
\sb
\sb
\sa
\sb
\sb
\sa
\sb
\sa
\sb
\sa
\sb
\sa
\sb
\sa
\sb
\sa

\noindent (published in Journal of Physics A: 
Mathematical and General {\bf 25} (1992) L157-L160)

\vfill\eject
\baselineskip = 0.84 true cm


\noindent The theory of quantum algebras (ie, quasi-triangular Hopf 
algebras or deformations of quantum universal enveloping 
algebras) and quantum groups (ie, compact matrix 
pseudo-groups), which originated with the works of 
Kulish and Reshetikhin [1], Drinfel'd 
[2], Jimbo [3] and Woronowicz [4] (among others), 
continues to attract the 
attention of both mathematicians and physicists. Up to now, most of the 
physical applications of quantum groups have been 
devoted to statistical mechanics, in connection with 
the solutions of the quantum Yang-Baxter equation, to solvable 
models, in connection with the theory of quantum inverse scattering, 
and to rational conformal field theory. Recently, there have 
been several applications in the domain of molecular 
and nuclear spectroscopy (mainly vibrational-rotational 
spectroscopy) [5-10]. 
Furthermore, attempts to apply quantum groups to atomic spectroscopy 
(viz, fine structure of the Hydrogen atom) [11] and to 
solid-state physics (viz, formation of coherent structures) [12] have 
been suggested very recently. 

\smallskip

It is the aim of this Letter to show that quantum groups can be 
also of interest in the field of chemical physics. More precisely, 
we want to show how the quantum algebra $su(2)_q$, one of the simplest
quantum algebras, can be used to derive an {\it Aufbau 
Prinzip} for atoms and monoatomic ions. 

\smallskip

It is well-known [13-21] that the atomic 
building principle for neutral atoms corresponds to the series 
$$
\eqalign{
& 1s \ll 2s < 2p \ll 3s < 3p \ll 4s < 3d < 4p \ll 5s < \cr
& 4d < 5p \ll 6s < 4f < 5d < 6p \ll 7s < 5f < 6d \cdots
}
\eqno (1)
$$
while the one for $N$-positive ions ($N$ $=$ degree of 
ionization) corresponds to the series 
$$
\eqalign{
& 1s < 2s < 2p < 3s < 3p < 3d < 4s < 4p < 4d < \cr
& 5s < 5p < 4f < 5d < 6s < 6p < 5f < 6d < 7s \cdots .
}
\eqno (2)
$$
(The sign $\ll$ in (1) serves to indicate rare gas.)
It is to be noted that among 99 neutral atoms, there are 20 
exceptions to the series (1) [19]. The series (1), often referred 
to as the Madelung-Klechkovskii series (cf.~Refs.~[13] and [14]), 
thus presents an 
approximate character. In contradistinction, the series (2) 
exhibits a more universal character since it has only 5 
exceptions for $N = 2$ and none for $3 \le N \le 6$ [19]. There is 
no model for describing in an unified way the series (1) and 
(2).  

\smallskip

The model we present here for the simultaneous 
description of (1) and (2) 
starts from the $O(4)$ symmetry of the Hydrogen atom. From the 
work of Fock [22], we know that the discrete spectral problem 
for the Hydrogen atom in $\grt$ is equivalent to that of a 
symmetrical rotor (or spherical top) in $\grq$. In the case of a 
many-electron neutral atom, the $O(4)$ symmetry is broken by 
inter-electronic repulsions and relativistic effects. In this 
respect, the chain of groups $SO(4) \supset SO(3)$ furnishes relatively 
good quantum numbers, viz, $n$ and $\ell$ (the principal and the 
orbital angular momentum quantum numbers). The model of Novaro 
[18] for neutral atoms relies on the chain $SO(4) \supset SO(3)$. 
In the latter model, the Hamiltonian $H$ spanning the 
$n\ell$ shells reads 
$$
H \, \simeq \, {1 \over {h + 1}} \qquad \quad h \, = \, {1 \over {2I}} 
( {\bf \Lambda}^2 + \alpha \, {\bf L}^2 )
\eqno (3)
$$
where $h$ is the Hamiltonian for an asymmetric rotor in $\grq$. 
In equation (3), ${\bf \Lambda}$ and ${\bf L}$ stand for the angular momenta in 
$\grq$ and $\grt$, respectively, and $\alpha$ is the asymmetry 
parameter given by $\alpha = I/I' -1$ in terms of the moments 
of inertia $I$ and $I'$. At this stage, it should be 
emphasized that the model based on (3) reproduces the series 
(1) in a reasonable way for $\alpha = 4/3$ [18] but that no admissible 
value of $\alpha$ reproduces the series (2). 

\smallskip

The basic ingredient of our model is to replace the chain 
$SO(4) \supset SO(3)$ by the $q$-deformed ``chain'' 
$SO(4) > SO(3)_q$. More precisely, we replace the 
constant $\alpha$ by a $q$-dependent parameter and the operator 
${\bf L}^2$ (whose eigenvalues are $\ell(\ell+1)$, with $\ell \in \grn$) 
by the Casimir operator of the quantum algebra $so(3)_q$. 
Therefore, the Hamiltonian $h$ is replaced by 
$$
h_q \, = \, {1 \over {2I}} \left( {\bf \Lambda}^2 + \alpha(q) \, 
[\ell]_q [\ell + 1]_q \right) 
\eqno (4)
$$
where the $[ \; ]_q$-integers are defined through
$$
\eqalign{
[x]_q \, & = \, {{q^x - q^{-x}} \over {q - q^{-1}}} \quad 
{\rm for} \quad x \in \grn \cr 
      \, & = \, q^{x-1} + q^{x-3} + \ldots + q^{-x+1} \quad {\rm for} \quad 
x \in \grn - \left\{ 0 \right\}.
}
\eqno (5)
$$  
Because ${\bf L}$ is the projection of ${\bf \Lambda}$ on a 
privileged axis, the $q$-deformation takes place along this 
axis. Thus, it seems natural to deform the inertial 
moment $I'$ (the one with respect to the privileged axis) 
without modification of the inertial moment $I$ (the one with 
respect to the three axes perpendicular to the privileged 
axis). Hence, we use the same $I$ in (3) and (4). The 
description of our Hamiltonian model $h_q$ is complete 
once the function $\alpha(q)$ is fixed. We shall adopt the 
linear law
$$
\alpha(q) \, = \, 3 - {5 \over 3} \, q. 
\eqno (6)
$$
Finally, we note that we can, of course, substitute for the 
operator ${\bf \Lambda}^2$ in (4), as well as in (3), its eigenvalues 
$\lambda(\lambda+2) \equiv (n-1)(n+1)$, with $\lambda \in \grn$.  

\smallskip

There are two interesting limiting cases for the 
model based on (4)-(6). The case $q=1$ (ie, $\alpha=4/3$) 
corresponds to the Novaro [18] model for neutral atoms 
and the case $q=9/5=1.8$ (ie, $\alpha = 0$) to the Hydrogen atom. 

\smallskip

Let us now examine the capabilities of our model. In this 
model, the negative energies of the various $nl$ shells are 
given by the eigenvalues of the Hamiltonian
$$
H_q \, = \, {E_0 \over {h_q + 1}}
\eqno (7)
$$
where $E_0$ is some arbitrary (negative) energy which reduces to the 
energy of the ground state of the Hydrogen atom for $q=9/5$. 
Following an usual practice in chemical physics and quantum 
chemistry, we note that the ordering afforded 
by (7) is the same as that obtained from
$$
\sqrt{\varepsilon_q(n,\ell)+1} \, = \, 
\sqrt{n^2 + \alpha(q) \, [\ell]_q \, [\ell + 1]_q}. 
\eqno (8)
$$
More precisely, we shall use
$$
\eqalign{
\varepsilon_q(n,s)+1 \, & = \, n^2 \cr
\varepsilon_q(n,p)+1 \, & = \, n^2 + (3 - {5 \over 3} q) (q + q^{-1}) \cr
\varepsilon_q(n,d)+1 \, & = \, n^2 + (3 - {5 \over 3} q) (q + q^{-1}) 
(q^2 + 1 + q^{-2}) \cr
\varepsilon_q(n,f)+1 \, & = \, n^2 + (3 - {5 \over 3} q) 
(q^2 + 1 + q^{-2}) (q^3 + q + q^{-1} + q^{-3}) 
}
\eqno (9)
$$ 
for ordering the energies of the orbitals $ns$, $np$, $nd$ and 
$nf$. 

\smallskip

We now discuss the results arising from equations (7)-(9). 
First, let us consider the case of positive ions. From formulas 
(9), we can see that the series (2) is reproduced for 
$1.15 \le q \le 1.30$. Therefore, the range $q=1.15$-$1.30$
is appropriate for $1 < N < 7$. Second, we note that for 
$1.6 \le q \le 1.8$, the order of the shells is hydrogenlike 
(in the sense of energy increasing with $n$). Such an order  
is convenient for highly-ionized atoms. Third, the case of 
neutral atoms is obtained for $q=0.85$. Indeed, the value 
$q=0.85$ reproduces the series (1) with a reasonable agreement~: 
for $n \le 6$, the agreement is perfect and for $n > 6$, the 
ordering (1) is respected with a deviation of less than $8\%$. 
This result reflects the fact that the Madelung-Klechkovskii 
rule presents several exceptions. 

\smallskip
 
In conclusion, we have derived a $q$-model ({\it Aufbau Prinzip}) which 
describes in an unified way (through equations (4)-(9)) 
neutral atoms, positive monoatomic 
ions, highly-ionized atoms and hydrogenlike ions. For 
neutral atoms, this model with $q=0.85$ gives a refinement of 
the Novaro model (which corresponds to $q=1$). The model for 
positively charged ions (corresponding to $q=1.15$-$1.30$) is 
entirely new. 
The application described in this Letter, 
which concerns the periodic structure of chemical 
elements, parallels the recent 
applications to nuclear [5,10], atomic [11], molecular [5-9] and 
solid-state [12] physics. Along the same vein (ie, the 
$q$-deformation of level splitting problems), we may reconsider 
the problem of mass formulas for nuclei and elementary 
particle physics. This problem shall be tackled in a 
forthcoming paper. Despite the present increase of pessimism 
concerning the quantum group invasion 
(possibly the quantum group pest), 
all these applications should invite one to pursue the 
investigations of quantum groups. 

\smallskip 

\noindent {\bf Acknowledgments}

\noindent One of the authors (M~K) thanks D~B Fairlie for his 
interesting comment about the future of quantum groups. The 
other author (T~N) is grateful to the {\it Institut de Physique 
Nucl\'eaire de Lyon} and the {\it Institut de Physique de l'Universit\'e 
d'Oran} for their help concerning his stay at Villeurbanne-Lyon.

\vfill\eject
  \baselineskip = 0.74 true cm

\noindent {\bf References}

\item{[1]} Kulish P~P and Reshetikhin N~Yu 1983 {\it 
J.~Soviet.~Math.} {\bf 23} 2435

\item{[2]} Drinfel'd V~G 1985 {\it Soviet.~Math.~Dokl.} {\bf 32} 254

\item{[3]} Jimbo M 1985 {\it Lett.~Math.~Phys.} {\bf 10} 63

\item{[4]} Woronowicz S~L 1987 
{\it Comm.~Math.~Phys.} {\bf 111} 613

\item{[5]} Iwao S 1990 {\it Progr.~Theor.~Phys.} {\bf 83} 363

\item{[6]} Chang Z and Yan H 1991 
{\it Phys.~Lett.} {\bf 154A} 254

\item{[7]} Chang Z, Guo H-Y and Yan H 1991 
{\it Phys.~Lett.} {\bf 156A} 192

\item{[8]} Chang Z and Yan H 1991 
{\it Phys.~Lett.} {\bf 158A} 242

\item{[9]} Bonatsos D, Argyres E~N and Raychev P~P 1991 
{\it J. Phys. A: Math. Gen.} {\bf 24} L403 

\item{[10]} Bonatsos D, Drenska S~B, Raychev P~P, Roussev R~P 
and Smirnov Yu~F 1991 {\it J.~Phys.~G: Nucl.~Part.~Phys} {\bf 17} L67

\item{[11]} Kibler M and N\'egadi T 1991 
{\it J.~Phys.~A: Math.~Gen.} {\bf 24} 5283

\item{[12]} Tuszy\'nski J~A and Kibler M 1991 
{\it J.~Phys.~A: Math.~Gen.} submitted 

\item{[13]} Madelung E 1936 {\it Die matematischen Hilfsmittel 
des Physikers} (Berlin: Springer)

\item{[14]} Klechkovskii V~M 1961 
{\it J.~Exp.~Theor.~Phys.} {\bf 41} 465 

\item{[15]} L\"owdin P~O 1969 {\it Int.~J.~Quantum Chem.} 
{\bf S3} 331

\item{[16]} Barut A~O 1971 in {\it Proc. Rutherford Centennial 
Symposium} ed B~G Wybourne (Christchurch: University of Canterbury 
Press)

\item{[17]} Rumer Yu~B and Fet A~I 1971 
{\it Teor.~i Mat.~Fiz.} {\bf 9} 203

\item{[18]} Novaro O 1973 {\it Int.~J.~Quantum Chem.} {\bf S7} 53

\item{[19]} Katriel J and J\o rgensen C~K 1982 
{\it Chem.~Phys.~Lett.} {\bf 87} 315

\item{[20]} Hefferlin R~A, Zhuvikin G~V, Caviness K~E and 
Duerksen P~J 1984 {\it J.~Quantum Spectrosc.~Radiat.~Transfer} 
{\bf 32} 257 

\item{[21]} Kibler M 1989 {\it J.~Molec.~Struct.~(Theochem)} 
{\bf 187} 83

\item{[22]} Fock V 1935 {\it Z.~Phys.} {\bf 98} 145 

\bye